# Cosmic Rays Self-arising Turbulence with Universal Spectrum −8/3

Artem S. Chefranov, Sergey G. Chefranov, and Georgy S. Golitsyn
Obukhov Institute of Atmospheric Physics RAS, Moscow, Russia
*Received 2022 April 22; revised 2023 May 11; accepted 2023 May 11; published 2023 July 3*

## Abstract

In the inertial subrange of scales, an exact compressible turbulence universal spectrum law $-8/3$ for the density fluctuations of cosmic rays (CRs) in the frame of the known two-fluid model of CR dynamics is obtained. It is shown that the origin of this scaling law may be due to the arising of shocks at the breaking of the nonlinear simple waves of CRs near the scale of their Larmor's radii, as it is well known for the solar wind with the same turbulent spectrum law $-8/3$. The consistency of the turbulence spectrum $-8/3$ of CRs with the observed nonthermal differential energy distribution of CRs with a similar index $-8/3$ due to the possibility of self-reacceleration of the CRs on the self-arising shocks is stated. The turbulent diffusion mechanism for the observed CRs energy spectrum breaks is considered.

*Unified Astronomy Thesaurus concepts:* Cosmic rays (329); Particle astrophysics (96); Interplanetary turbulence (830); Ultra-high-energy cosmic radiation (1733); Spectral energy distribution (2129); Interplanetary shocks (829)

## 1. Introduction

Cosmic rays (CRs) are the most ubiquitous population of charged relativistic particles in the universe, playing a crucial role in understanding the problems of high-energy physics and plasma physics, as well as the fundamental and applied problems of atmospheric electricity and climate variability on Earth (Cesarsky 1980; Blandford & Eichler 1987; Golitsyn 1997, 2005, 2012, 2021; Malkov & O'C Drury 2001; Carslaw et al. 2002; Adriani et al. 2011; Mertsch & Sarkar 2014; Grinier et al. 2015; Aguilar et al. 2020; Niu 2021).

CRs are particles with energies $E > 1\,\mathrm{GeV\,nucleon^{-1}}$ pervading space, having a nonthermal energy distribution typically of a power-law spectrum rather than a Maxwell–Boltzmann form (Cesarsky 1980; Carslaw et al. 2002; Adriani et al. 2011; Grinier et al. 2015; Aguilar et al. 2020; Niu 2021). The differential energy spectrum $I(E) = dN(E)/dE$ according to the observational data (see Figure 1 below and in Liu et al. 2022) for protons (which make up the main part of CRs) with energies $200\,\mathrm{GeV} < E < 3 \times 10^6\,\mathrm{GeV}$ has a power-law dependence $I(E) \propto E^{-2.67\pm0.02}$ (Adriani et al. 2011; Grinier et al. 2015; Aguilar et al. 2020; Niu 2021). Moreover, as represented in Figure 1, the power-law energy spectrum $-8/3$ is observed up to energies $E \approx 10^{11}\,\mathrm{GeV}$ except for the subrange of energies between the so-called "knee" ($E \approx 3 \times 10^6\,\mathrm{GeV}$) and "ankle" ($E \approx 10^9\,\mathrm{GeV}$) regions.

To date, the problem of determining the origin of CR particles and the corresponding mechanism for the formation of the observed differential energy spectrum, represented in Figure 1 and characterized by a power law close to $-8/3$, remains mysterious (Adriani et al. 2011; Grinier et al. 2015; Aguilar et al. 2020; Niu 2021).

One of the conventional mechanisms known for the formation of this spectrum is associated with the repeated passage of particles through the shock wave front where the main acceleration site of CRs is obtained (Blandford & Eichler 1987; Malkov & O'C Drury 2001; Mertsch & Sarkar 2014). It is generally believed that collisionless shock waves in supernova remnants (SNRs) are the dominant agent for acceleration. The hypothesis that supernovae are the most likely energy source of CRs has been strengthened, and significant advances have been made in understanding particle acceleration in supernova remnant environments (Reichherzer et al. 2022). After the shock cannot contain particles, anymore they diffuse through the interstellar medium (ISM), producing secondary particles. The CR particles distribution function at the shock is roughly a power law (Blandford & Eichler 1987; Malkov & O'C Drury 2001; Mertsch & Sarkar 2014):

$$I(E) \propto E^{-\gamma};\quad \gamma = s - 2;\quad s = 3\kappa/(\kappa - 1);\quad \kappa = \rho/\rho_0, \quad (1)$$

where $\kappa$-compression and $\rho; \rho_0$ is the matter density of ISM behind and before the shock wave front. Thus, the observed value $\gamma \approx 8/3$ needs the value of compression $\kappa = 14/5 = 2.8$. When $\kappa = 4$ from Equation (1) it is possible to obtain the value $\gamma = 2$ (Mertsch & Sarkar 2014).

Moreover, from the theory of similarity (Golitsyn 1997, 2005, 2012, 2021) derives the same exponent $\gamma = 8/3$.

It is important to distinguish the number density of particles $\tilde{n}(E) \propto \tilde{n}(p) \propto p^{-8/3}$, having a fixed energy $E \approx pc$ or momentum $p$ from the total number density of particles $n(\vec{x};\ t) = \int d^3p f(t;\ \vec{x};\ \vec{p})$, which is independent of the magnitude of the particle momentum. The measured quantity is the particle flux from one hemisphere of the sky $F(E) = \pi I(E)$, where $I(E)$ is the number of particles with energy $E$ from one steradian per unit time through the unit area when the particles number density is $\tilde{n}(E) = 4\pi I(E)/V(E) 4\pi I(E)/c$, where $V(E)$ and $c$ velocity of CR particles and the velocity of light in a vacuum (Golitsyn 1997, 2021).

Indeed, as it is known, the distribution function $f(t;\ \vec{x};\ \vec{p})$ that satisfies a certain kinetic equation can be used to describe CR particles, which determines $\Delta\tilde{n} = f(t;\ \vec{x};\ \vec{p})\Delta\mu;\ \ \Delta\mu = \Delta\vec{x}\Delta\vec{p}$ the number of particles contained in an element of the phase volume $\Delta\mu$. At the same time, taking into account the ratio $I(E) \propto p^2 f$ given in Blandford & Eichler (1987), the representation $\tilde{n}(t;\ x;\ p) \propto 4\pi p^2 f/c$ is valid when the dependence $\tilde{n}(p) \propto p^{-8/3}$ corresponds to the observed scaling law $f(t;\ \vec{x};\ \vec{p}) \propto p^{-14/3}$.

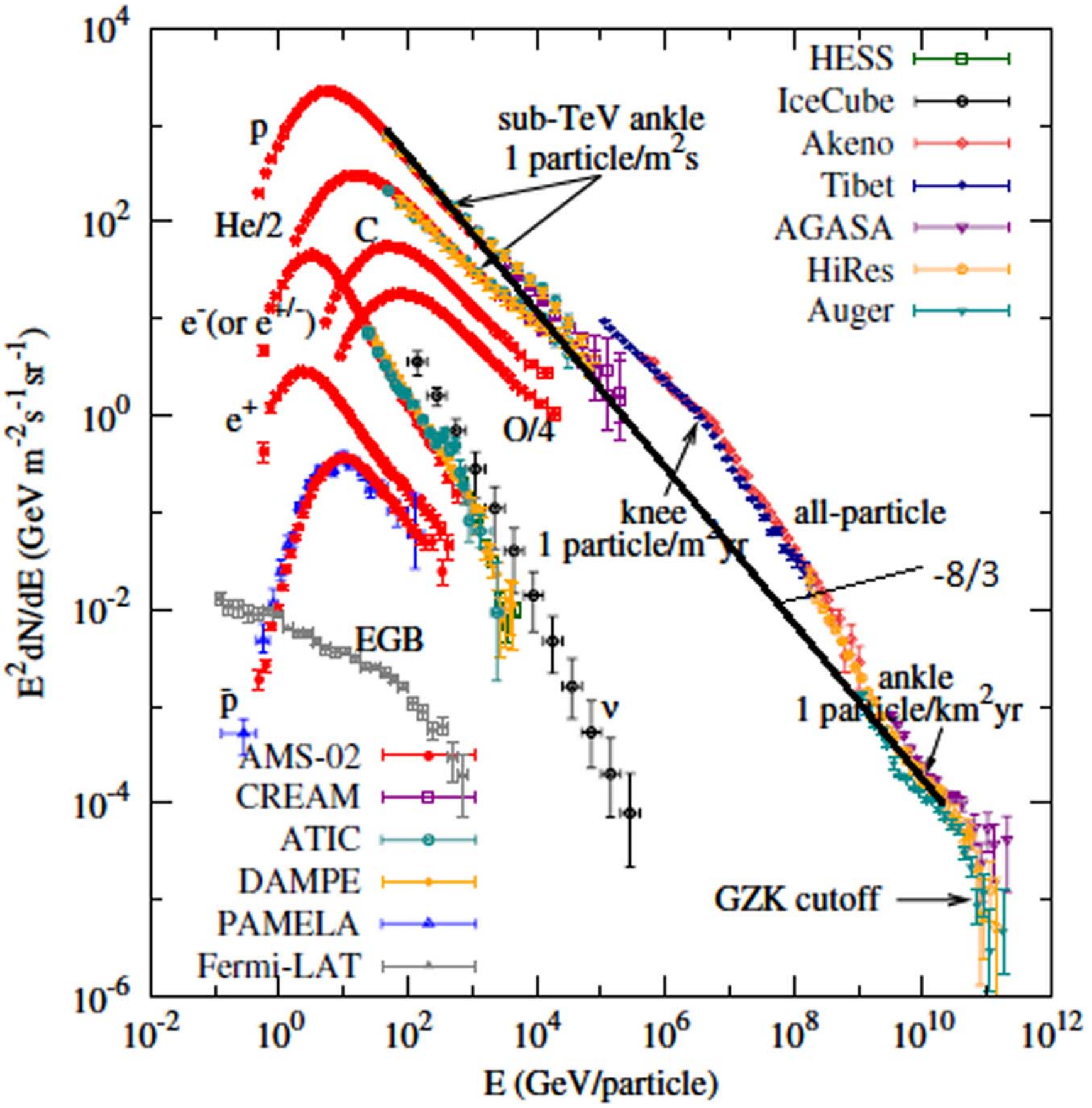

**Figure 1.** Observational data, where the black solid line represents the energy power-law spectrum with an exponent $-8/3$, added to Figure 1 of Liu et al. (2022).

Therefore, the problem of understanding the mechanism leading to the establishment of the observed scaling law $\tilde{n}(p) \propto p^{-8/3}$ for the specific density of particles turns out to be closely related to the mechanism of formation of the general form of the distribution function $f(t;\ \vec{x};\ \vec{p})$. The latter gives the number density $n(t;\ \vec{x})$ of particles, as well as its two-point correlation function and the corresponding turbulent spectrum. An analog of such consideration was carried out by Armstrong et al. (1995). Based on the obtaining of a two-point correlation function from the observational data and the corresponding turbulent spectrum for fluctuations in the electron distribution density in the ISM, a scaling similar to the Kolmogorov–Obukhov law in the inertial scale interval was established (see Figure 2 and Figure 1 in Armstrong et al. 1995).

Thus, the question of whether there is a correlation between the observed differential energy spectrum of CRs and the turbulent spectra characterizing the shock waves themselves has not been widely considered. For example, in Chefranov & Chefranov (2021), for the inertial subrange, a universal spectrum of strong turbulence with an exponent $-8/3$ is obtained, corresponding to the time of occurrence of the shock wave. In particular, this made it possible to explain the mechanism of the formation of the $-8/3$ turbulent spectrum observed in the cosmic plasma of the solar wind and the magnetosphere for the density and magnetic field fluctuations (Passot & Sulem 2015; David & Galtier 2019) and adequacy of the $-8/3$ energy spectrum for the Kadomtsev–Petviashvili theory of strong acoustic turbulence (Kadomtsev & Petviashvili 1973; L'vov et al. 1997; Kuznetsov 2004) is also established in Chefranov & Chefranov (2021).

In our paper, we obtain the two-point correlation function, and the corresponding universal turbulence spectrum law $-8/3$ is obtained for the CR number density $n(\vec{x};\ t)$ fluctuations. It is done on the basis of an exact analytical solution to the problem of compressible turbulence (Chefranov & Chefranov 2021) by using a specific solution of the Boltzmann–Vlasov equations with $f(t;\ \vec{x};\ \vec{p}) \propto n(\vec{x}, t)p^{-5}$, which is obtained within the frame of the theory (Skilling 1971, 1975). In the case of a balance between Fermi acceleration and dissipation of the CR particle energy, the isothermal mode of propagation of CR particles is obtained in Skilling (1975) and the motion of a medium consisting of such particles can be described by Euler hydrodynamics equations. Indeed, in the two-fluid theory (Malkov & O'C Drury 2001), based on kinetic equations, an analog of the one-dimensional Euler equation is also derived to describe a compressible polytropic medium consisting of CR particles (see Equations (6.4) and (6.5) in Malkov & O'C Drury 2001).

This paper considers the possibility of self-generation of turbulence with an energy spectrum having an index of $-8/3$ in a compressible medium consisting only of CR particles. In a certain range of scales of the inertial interval, the turbulent energy spectrum with an index of $-8/3$, as can be seen from Figure 2, describes the observational data more adequately than the Kolmogorov–Obukhov indicator $-5/3$, which is usually considered for the turbulent energy spectrum (Reichherzer et al. 2022).

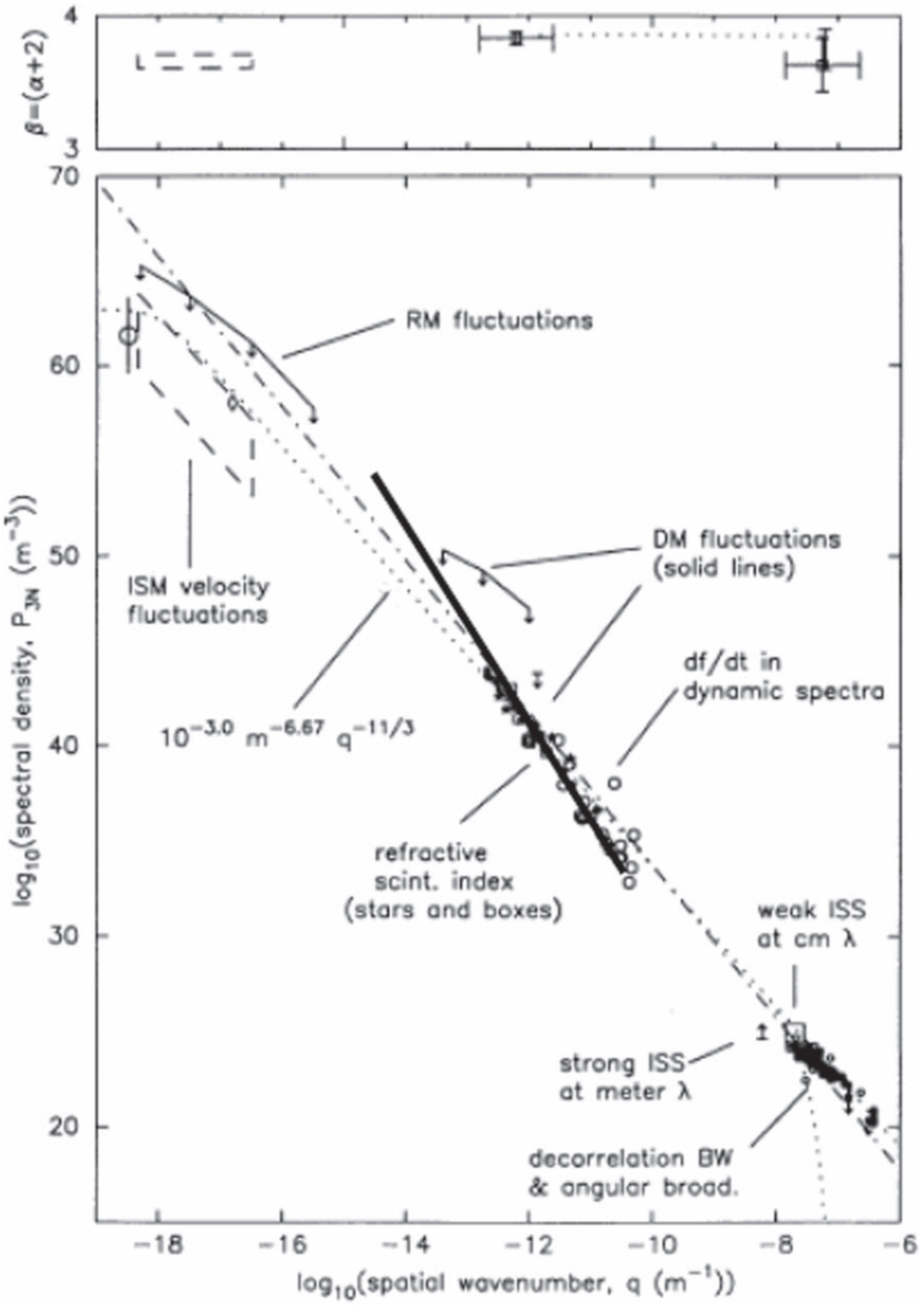

**Figure 2.** The turbulent spectrum of electron density fluctuations with an exponent $-11/3$, following from the Kolmogorov–Obukhov law $-5/3$, is represented by dots. The slope corresponding to the energy spectrum of $-8/3$ gives an exponent $-14/3$ (see the solid black line added to Figure 1 of the article Armstrong et al. 1995).

The mechanism of such self-generation of turbulence is associated with the collapse of simple waves and the corresponding occurrence of shock waves in a finite time determined by the magnitude of the spatial velocity gradient of CR particles (see Equation (11)). At the same time—in contrast to the turbulence previously considered, initiated by fast CR particles in the ISM via stream instability (Blasi et al. 2012)—in this case, turbulence occurs precisely in an environment consisting only of CR particles themselves. As a result, a new mechanism for the reacceleration of CR particles in such self-generated shock waves in a medium consisting of CR particles is also obtained. It not only complements but also expands the applicability of the implementation of the reacceleration of CR particles, compared with the previously known mechanism of reacceleration in relatively weak shock waves at a great distance from the source of the primary acceleration (Thoudam & Horandel 2014; Blasi 2017). Additionally, in this work, a new dependence of the CR diffusion coefficient $D(E) \propto E^{\delta}$ on their energy $E$, corresponding to the $-8/3$ spectrum, with a different index $\delta = -2/3$ from those, previously considered, is obtained. Indeed, there is a standard relation (Erlykin & Wolfendale 1998; see also Jokipii 1966; Giacalone & Jokipii 1999) for the diffusion coefficient $D(k) \propto \lambda c / k E_T(k)$, as a function of the parameters of the turbulent medium. Here $\lambda$ is the Larmor radius of the particle of energy $E$ and $E_T(k)$ is the power-law turbulent spectrum of the irregularities of the field on the scale $k = 2\pi/\lambda \propto 1/E$. Thus, a known relation for an exponent $\delta = 2 - \chi$ in the relation $D(E) \propto E^{\delta}$ is valid for the power-law turbulent spectrum $E_T(k) \cong k^{-\chi}$. For example, in the case of turbulence usually considered in this connection, which corresponds to the Kolmogorov–Obukhov energy spectrum with $\chi = 5/3$, the value of this index is positive and equal to $\delta = 1/3$, but for the energy spectrum with $\chi = 8/3$ it is obtained new estimation of index $\delta = -2/3$.

The use of a combination of the new ideas obtained in this paper regarding the possible mechanism of reacceleration, along with the dependence of the diffusion coefficient in the form $D(E) \propto E^{-2/3}$ is considered in connection with the problem of the observed spectral anomaly at energies nearly 100 GeV (Blasi et al. 2012; Neronov et al. 2012; Niu 2021; Reichherzer et al. 2022). In particular, a correspondence was obtained with the observational data for the ratio of the flow of antiprotons to the flow of protons, which turns out to be generally independent of energy starting with an energy of about 60 GeV (Aguilar et al. 2016). These data cannot be explained within the framework of the traditional approach, in which the ratio of the flow of secondary particles to the flow of primary particles is assumed to be proportional to the energy-dependent value in the form of $D^{-1}(E) \propto E^{-\delta}$.

The next Section 2 provides a description of the conditions for the implementation of the theory (Skilling 1975). In Section 3, based on the theory (Chefranov & Chefranov 2021), a turbulent spectrum of fluctuations in the density of CR particles with a universal exponent $-8/3$ is obtained. Section 4 discusses the comparison of the conclusions of the theory and observational data, as well as the results of the similarity theory (Golitsyn 1997, 2005, 2012, 2021). This section also analyzes the effect of turbulent diffusion on the formation of the observed proton energy spectrum.

As an alternative to the known explanations of the observed correlation between gamma-ray bursts and CRs (Waxman 1995; Dado & Dar 2012; Gill & Granot 2022), it is proposed to consider the general problem of the origin of CRs on the basis of a relatively new cosmological theory of the accelerated expansion of the universe without the Big Bang (Chefranov & Novikov 2010; Novikov & Chefranov 2011; Mirzoeva & Chefranov 2018). At the same time, gamma rays and CRs may have a common mechanism of origin due to the continuous birth of primary ultralight (having mass $\propto 3 \times 10^{-33}$ eV) from the vacuum scalar bosons, when the corresponding polarization of the vacuum occurs during the expansion of the universe. At the same time, the mechanism of transformation of such bosons into emerging photon pairs may be similar to the pion decay considered in Hayakawa (1952), and be responsible for the observed difference in the intensity of the diffuse gamma-ray flux near the central regions of the Galaxy and on its periphery (Amellino-Camelia 1998; Acero et al. 2016). In the case of such a mechanism of origin of both diffuse and explosive gamma-ray fluxes, this difference in different regions of the Galaxy may indicate the spatial heterogeneity of such a process of boson generation from the vacuum.

## 2. The Inertial Motion of Cosmic Ray Particles

The Skilling (1971) analysis is based on the collisionless Boltzmann–Vlasov equation, which expresses the conservation of CR particles in phase space:

$$\frac{\partial f}{\partial t} + \frac{\partial}{\partial x_i}(fV_i) + \frac{\partial}{\partial p_i}\left(f\frac{dp_i}{dt}\right) = 0; \quad i = 1, 2, 3. \tag{2}$$

Here and further, on the repeating indexes, the summation from 1 to 3 is meant.

Since protons are the dominant CR species, it is taking $f$as referring to protons. In Skilling (1971, 1975) Equation (2) is transformed to the so-called "wave frame" which gives the possibility to take into account the interactions between CR particles and Alfvén waves.

In Skilling (1975) the possibility of the balance between the acceleration of charged particles of CRs in random magnetic fields and dissipative factors is established.

The formation of the observed spectrum of CR energy can be influenced by factors that determine their spatial transfer process, including turbulent diffusion in turbulence, which is caused by CRs themselves. However, when considering the mechanism of formation of such a turbulent regime, it is possible to use an approximate approach developed in the mentioned work (Skilling 1975), in which the initial transfer processes are described in a simplified way. Indeed, for the ultrarelativistic particles, in the approach when in Skilling (1975) is considered the case of negligible background plasma motion $V_0 \ll V_A$ (where $V_0$-is the velocity of background plasma, $V_A$-is the Alfvén velocity) when $f(t;\ \vec{x};\ \vec{p}) \propto n^a(\vec{x}, t)p^{-s};\ s > 4, a = s/(2s - 5)$.

Thus in this approach of Skilling (1975) the transfer of CR moving at a speed close to the speed of light $c$ for the transfer time $t \gg t_D = D_0/c^2 \approx 1$ yr–10 yr, when the characteristic values of the diffusion coefficient are $D_1 0 \approx 0^{28}$ cm$^2$ s$^{-1}$–$10^{29}$ cm$^2$ s$^{-1}$ (Kempski & Quataert 2022) is considered. See also the same estimations in Hayakawa (1952) giving the near ballistic transfer of CRs.

Here $n$—is the number density of CR particles in their convection with the Alfvén wave. As a result, the equation for the CR particle momentum is obtained (see Equations (26) and (27) in Skilling 1975):

$$\frac{1}{p}\frac{dp}{dt} = -b\frac{\partial u_i}{\partial x_i};\ p \propto n^b b = \frac{1}{3}\left[1 - \frac{2s(s-4)}{(2s-5)(s-3)}\right]. \tag{3}$$

In Equation (3) $\vec{u} \propto V_A \vec{n_H}$, where $\vec{n_H}$-is the unit vector directed along the background magnetic field. In Skilling (1975) the CR particles are considered as a polytropic gas with pressure $P_{CR} \propto p^5 f \propto n^{b(5-s)+a}$, number density $n_{CR} \propto p^3 f \propto n^{a+b(3-s)}$, and adiabatic exponent:

$$\gamma = \frac{d \ln P_{CR}}{d \ln n_{CR}} = 1 + \frac{2(5-s)}{(s-3)(2s-5)}. \tag{4}$$

From Equations (3) and (4) it is possible to obtain the condition for inertia motion of CR particles with invariant energy and momentum $p$ = const when the net energy loss from CRs is small (Skilling 1975):

$$s = 5;\ b = 0;\ a = 1, \tag{5}$$

$$\gamma = 1. \tag{6}$$

In this case, when conditions are valid the distribution function has representation $f(t;\ \vec{x};\ \vec{p}) \propto n(\vec{x};\ t)p^{-5}$, which gives a scaling law $N(t;\ \vec{x};\ \vec{p}) \cong 4\pi n(\vec{x};\ t)/cp^3$ that is near the same as a known scaling law for the differential energy spectrum with $N \propto p^{-8/3}$.

In addition, when the conditions in Equations (5) and (6) are met, the Euler equation for the velocity field of such a medium, the continuity equation for density, and the isothermal equation of state, respectively, can be used to describe the isothermal mode of motion of a medium consisting of CR particles:

$$\frac{\partial V_i}{\partial t} + V_j \frac{\partial V_i}{\partial x_j} = -\frac{1}{\rho}\frac{\partial P_{\mathrm{CR}}}{\partial x_i};\; i = 1,.., d, \tag{7}$$

$$\frac{\partial \rho}{\partial t} + \frac{\partial}{\partial x_i}(\rho V_i) = 0;\; \rho = m_p n_{\mathrm{CR}}, \tag{8}$$

$$P_{\mathrm{CR}}\rho^{-\gamma} = \mathrm{const};\; \gamma = 1;\; P_{\mathrm{CR}} = c_S^2 \rho. \tag{9}$$

In Equations (7) and (8) by repeating indices, the summation is implied, where $d$ is the dimension of the space. In Equation (9) $c_S$-is the constant isothermal sound speed of such exotic medium, consisting of CR particles. Equations (7) and (8) exactly coincide with the equations of the two-fluid model (see Equations 6.4 and 6.5 in Malkov & O'C Drury 2001) under the conditions of conservation of CR energy, according to the theory (Skilling 1975). At the same time, the more general system of equations considered in Malkov & O'C Drury (2001) admits decoupling and allows the description of a medium of CR particles based on Equations (7)–(9), regardless of the equations describing the motion of the medium in which CRs propagate.

For arbitrary finite Mach numbers, the exact solution of the system (Equations (7) and (8)) was obtained by Riemann in 1860 only for the one-dimensional case and in the form of an implicit dependence of a nonlinear simple wave on the initial velocity and density fields (Landau & Lifshitz 1987). The peculiarity of a simple wave is that it has a functional relation between the density of the medium and the wave velocity.

For a medium consisting of CR particles, however, it turns out to be characteristic of the special case discussed below, when the adiabatic index in Equation (9) is equal to one. For the one-dimensional case $d = 1$, the exact solution of the system (Equations (7)–(9)) in the form of a nonlinear simple Riemann wave taking into account (Chefranov & Chefranov 2021) has the explicit form:

$$\begin{aligned} \rho(x, t) &= \int_{-\infty}^{\infty} d\xi \rho_0(\xi)\left(1 + t\frac{dV_0}{d\xi}\right) \\ &\times \delta(\xi - x + (\pm c_S + V_0(\xi))t); \\ \rho_0(x) &= \rho(x, t = 0) = \rho_{00}\exp\left(\frac{V_0(x)}{c_S}\right); \\ V_0(x) &= V(x;\; t = 0). \end{aligned} \tag{10}$$

In Equation (10) $\delta$-is the Dirac delta function, $\rho_0(x)$, $V_0(x)$—are the initial density and velocity fields.

The solution for the velocity field has a similar form to Equation (10), in which it only needs to be replaced by $\rho_0(\xi) \to V_0(\xi)$ in Equation (10). In Equation (10), the value $\rho_{00} = \mathrm{const}$ is determined taking into account the relationship of the initial conditions for the density of the medium and for the distribution of the velocity field in it, which is characteristic, precisely when considering a nonlinear simple Riemann wave.

Note that the solution in Equation (10) of the one-dimensional system of Equations (7)–(9) preserves smoothness only over a finite time interval $0 \leqslant t < t_0$, where the collapse time of a simple Riemann wave or in other words the shock wave arising time is explicitly defined in Chefranov & Chefranov (2021) and depends only on the initial distribution of the velocity field and the adiabatic exponent (see Equation (33) in Chefranov & Chefranov 2021) $t_0 = \frac{2}{(\gamma+1)\left(\max_{x=x_M}\left|\frac{dV_0}{dx}\right|\right)}$.

For the case considered in Equation (9), where $\gamma = 1$ this minimal existence time of a smooth solution has the form

$$t_0 = \frac{1}{\left(\max_{x=x_M}\left|\frac{dV_0}{dx}\right|\right)}. \tag{11}$$

In Equation (11) $x = x_M$, this is the place where the nonlinear collapse of a simple Riemann wave occurs for the first time.

In Equation (11), this can be done for any smooth initial velocity field having a finite value of the energy integral in an unbounded space. For example, for the initial velocity field of the form $V_0(x) = a\exp(-x^2/2x_0^2)$ from Equation (11), we obtain an estimate $t_0 = x_0/V_0(x = x_0)$; $V_0(x_0) = a/\sqrt{e}$, for the minimum collapse time (loss of smoothness or collapse of a nonlinear wave) of the solution, which occurs for the first time at the value $x_M = x_0$. We assume, according to Wentzel (1974), for the drift velocity (in the coordinate system associated with the MHD wave, as in Skilling 1971, 1975) of CR particles in the Galaxy, the magnitude is $a \propto 36$ km s$^{-1}$ and the corresponding characteristic scale of spatial inhomogeneity is of the order of $x_0 \approx 3.085 \times 10^8$ km. Then, according to Equation (11), we obtain an estimate $t_0 \approx 1.4 \times 10^7$ s $\approx 0.5$ yr for the collapse time of a nonlinear wave describing a medium consisting of CR particles. The characteristic scale of spatial inhomogeneity used to estimate the collapse time of a nonlinear wave has a value close to the Larmor radius of protons when they have an energy of about 100–200 GeV (Grinier et al. 2015). In turn, it is at such energies that the spectral law $-8/3$ is established in the observational data (Adriani et al. 2011; Grinier et al. 2015). A similar change in the index of the turbulence spectrum in the solar wind plasma and the establishment of the spectral law $-8/3$ also takes place, starting from the scale of the Larmor radius of ions. Therefore, it can be assumed that, as in the cosmic plasma of the solar wind, the establishment of the spectral law $-8/3$ is due to the mechanism noted in David & Galtier (2019) associated with the collapse of a nonlinear simple Riemann wave in the medium of CR particles. This mechanism can be considered as an alternative to the existence of additional unknown sources of primary CRs and other variants noted in Grinier et al. (2015) in connection with the observation of the spectrum fracture and the establishment of the $-8/3$ law.

Note also that Equation (7), for the motion of particles by inertia (see also Chefranov 1991 and Chefranov & Chefranov 2019, 2020) with zero resultant force, according to the $d$-dimensional Hopf equation, exactly coincides with the Boltzmann–Vlasov, Equation (2), for the case of zero force $d\vec{p}/dt = 0$ (Skilling 1975) and with the generalized Liouville equation in the statistical mechanics of non-Hamiltonian systems (Ezra 2004).

## 3. Exact Turbulence Scaling Law −8/3 for the Cosmic Ray Density Fluctuations

Based on the exact solution of the Euler Equations (7)–(9) in the form of Equation (10), we obtain representations for the two-point correlation function of density $R_\rho(r) = \frac{1}{L\rho_{00}^2}\int_{-\infty}^{\infty} dx \rho(x + r;\; t)\rho(x;\; t)$ and the corresponding spectrum $E_\rho(k) = \frac{1}{2\pi}\int_{-\infty}^{\infty} dr R(r)\exp(-ikr)$ of density fluctuations in a medium consisting of CR particles. Here is used the external integral scale $L$, where $L = P_0^2/E_0$; $P_0 =$

$\int_{-\infty}^{\infty} dx|V_0(x)|$; $E_0 = \int_{-\infty}^{\infty} dx V_0^2(x)$ (Chefranov & Chefranov 2021). Thus, to describe the turbulence spectrum here, as in Chefranov & Chefranov (2021), an averaging over a spatial variable in the unbounded case is used.

In the limit $k \gg L^{-1}$ it is possible to obtain asymptotic representation (Chefranov & Chefranov 2021):

$$E_\rho(k) = C_\rho k^{-8/3} + O(k^{-10/3})$$
$$C_\rho = \frac{2^{1/3}}{c_S^2 L}\left(\frac{dV_0}{dx}\right)_{x=x_M}^{8/3}\left(\frac{d^3V_0}{dx^3}\right)_{x=x_M}^{-2/3} \times \Phi^2(0)\exp\left(\pm\frac{2V_0(x_M)}{c_S}\right). \quad (12)$$

In Equation (12) $\Phi(z) = \sqrt{\pi} Ai(z)$ is the Airy function, where $\Phi(0) = \frac{\sqrt{\pi}}{3^{2/3}\Gamma(2/3)} \approx 0.629$. The representation (Equation (12)) is obtained by using the method of stationary phase in the estimation of an integral $E_\rho(k)$ at the time $t = t_0$, where $t_0$ is determined in Equation (11). So the main asymptotic term in Equation (12) is determined with precise of the next term which has an order $O(k^{-10/3})$ in the limit of large wavenumber $k \gg 1/L$.

The representation of the energy spectrum can be obtained in a similar way in the form of a Fourier transform from the correlation function of the velocity field. Thus, the turbulence energy spectrum coincides with Equation (12) if in it the next replacement $C_\rho \to C_E$, $C_E = C_\rho c_S^2 \exp(\mp 2V_0(x_M)/c_S)$ is provided (Chefranov & Chefranov 2021).

The universality of the spectrum (Equation (12)) is manifested in the independence of the exponent $-8/3$ from the type of the initial velocity field.

Thus, in a medium consisting of CR particles, regardless of the initial conditions, a turbulent spectrum of density fluctuations corresponding to the universal scaling law $-8/3$ should be observed in the inertial subrange of scales. Indeed, according to the above estimate, the collapse time of a nonlinear wave in such an environment has a value of about 6 months. Therefore, such regularly occurring collapses seem to be realized continuously for an unlimited time, but always each such collapse of a nonlinear wave leads only to the marked spectral law $-8/3$ in Equation (12), despite the difference in the constant factor $C_\rho$ in Equation (12), which is depended on the initial conditions.

## 4. Discussion and Comparison with Observational Data

The exact solution (Equation (12)) for the universal turbulent spectrum of fluctuations in the density of CR particles is characterized by the scaling law $-8/3$ in the inertial subrange of scales, which does not depend on the initial conditions and on any parameters characterizing the equation of state of the medium. At the same time, the multiplier $C_\rho$ does not have such a universality property and significantly depends on the initial distribution of the velocity field $V_0$ and its derivatives at the point $x = x_M$, where, for the first time at the time found via Equation (11), a nonlinear simple wave describing the density distribution of CR particles collapsed. In addition, in Equation (12) there is a significant dependence of this multiplier on the constant value of the speed of sound $c_S$, characterizing the equation of state of a medium consisting of CR particles according to Equation (9). In this regard, for the possibility of the most complete comparison with the obtained exact solution (Equation (12)), it is of interest in future experiments to consider the data of observations of fluctuations in the density of CR particles carried out simultaneously at two different points in space along the direction of propagation of CR particles. At the same time, data on measuring the magnitude of the speed of sound $c_S$ in a medium consisting of CR particles would be of particular importance.

It should be noted that at present there are only measurement data of the dependence of the density of CR particles on their energy related to observations carried out on board only one spacecraft (Adriani et al. 2011; Grinier et al. 2015). The dependence of the proton density in CRs on their energy obtained in observations has a form $N(E) \propto E^{-2.67\pm0.02}$ for protons with energies $10\,\mathrm{GeV} < E < 3 \times 10^6\,\mathrm{GeV}$ that is consistent with a similar estimate of the scaling index $-8/3$, obtained from the theory of dimension and similarity in Golitsyn (1997, 2005, 2012, 2021). The corresponding similarity considerations are stated in the introduction of two parameters: the CR energy generation rate $G_{\mathrm{CR}} \propto 1 - 5 \times 10^{40}\,\mathrm{erg\,s^{-1}}$ and the energy volume densities of CR $w_{\mathrm{CR}} \propto 10^{-12}\,\mathrm{erg\,cm^{-3}}$ (see Equation (13) in Golitsyn 1997).

If we use these two parameters, supplementing them with such a dimensional parameter as the speed of light $c$ in a vacuum, then, for dimensional reasons, we can obtain an analog of scaling's law (Equation (12)) in the form $E_\rho \propto \left(\frac{w_{\mathrm{CR}} c}{G_{\mathrm{CR}}}\right)^{5/6} k^{-8/3}$.

It should be noted that the scope of applicability of the universal scaling law (Equation (12)) considered above, characterized by the indicator $-8/3$, is limited only by the inertial range of scales, in which there is no influence of both the scale of the source or pumping of turbulence energy and the scales at which the energy dissipation of turbulent fluctuations is realized. Therefore, the application of Equation (12) in connection with the analysis of a similar power law $-8/3$ depending on the differential spectrum of the energy of CR particles on the magnitude of the energy of these particles also applies only to a limited range of energies $10\,\mathrm{GeV} < E < 3 \times 10^6\,\mathrm{GeV}$.

In addition, according to the available observational data (Blasi et al. 2012; Neronov et al. 2012), the scaling law with an index of $-8/3$ is not implemented in the entire energy interval $10\,\mathrm{GeV} < E < 3 \times 10^6\,\mathrm{GeV}$, discussed above, which can be considered as an analog of the inertial scale interval in the theory of turbulence.

Indeed, according to observations, it turns out that in the energy range $10\,\mathrm{GeV} = E = 200\,\mathrm{GeV}$, the exponent in the scaling law, although not significantly, still exceeds the value of $8/3$ and is approximately equal to 2.85 (or even 2.9–3.1; Blasi et al. 2012; Neronov et al. 2012).

At the same time, the question arises as to how justified the application of the theory developed in this work can be to the specified range of energy values. This question needs to be answered in the context of the general problem of the origin and acceleration of CR particles.

Although the old idea that SNR acceleration is responsible for CRs of energy, and indeed up to the knee region, is supported by everyone, it is true that up to now direct evidence of it has been lacking (Erlykin & Wolfendale 2002). Fluctuations in the proton spectra are dependent on the geometry of the SNR in the vicinity, when the injected power is proportional to the density of the SNR. The situation in the halo is even more complex because there is no such source of energy as SNR explosions except as some shocks come out into this region.

Thus, the main source of energy and turbulence is presumable CRs escaping from the galactic disk which supports the turbulence in the halo and eventually establishes an equilibrium between CRs and magnetic fields. We will further present the arguments presented in Erlykin & Wolfendale (2002) for the possibility of establishing a relationship between the exponent $\gamma_{\rm ing}$ of the differential spectrum of the energy $dN(E)/dE \propto E^{-\gamma_{\rm ing}}$ of CR particles injected from the SNR with the exponent of the degree of turbulence energy in general and in particular with the exponent $-8/3$ in Equation (12). Therefore, in the halo, CRs themselves are the source of energy for turbulent motions. The mechanism of magnetic turbulence excited by the charged particles has the same resonant character as the mechanism of the scattering on magnetic irregularities. Hence the energy spectrum of protons $dN(\lambda)/d\lambda \propto \lambda^{-\gamma_{\rm ing}}$ creates a spectrum of sizes with the same slope as the differential energy spectrum $dN(E)/dE \propto E^{-\gamma_{\rm ing}}$, where $\lambda$ ($\lambda = p_\perp \big/ qB$; $p_\perp = mV_\perp \Big/ \sqrt{1 - V_\perp^2/c^2}$) is the Larmor radius, which at relativistic energies $E \approx p_\perp c; p_\perp \gg m_p c$ is proportional to the proton energy $\lambda \propto E$.

On the other hand, for the energy spectrum usually considered in the theory of turbulence, it is possible to introduce a definition in the form $E_T(k) = c\lambda qB\frac{d\lambda}{dk}dN(\lambda)\big/d\lambda \propto k^{-\chi}$, when, for example, $\chi = 5/3$ for the Kolmogorov spectrum, and $\chi = 8/3$ for the spectrum found in Equation (12). Taking into account the relationship between the wavelength $\lambda$ and the wavenumber $k = 2\pi/\lambda$, a known ratio between the exponents is obtained (Erlykin & Wolfendale 2002):

$$\chi = 3 - \gamma_{\rm ing}. \tag{13}$$

Equation (13) provides a relationship between the exponent $\chi$, of the spectrum of turbulence energy, the stationary mode of which is realized only in a certain inertial interval of scales, with the value $\gamma_{\rm ing}$, characterizing the energy source of particles injected at scales corresponding to the upper boundary of the inertial interval. According to Erlykin & Wolfendale (1998), the inertial scale interval for turbulence in a medium consisting of CR particles is limited from above by a value of 100 pc (corresponds to energy $E \approx 10$ GeV), and from below by a value of 1 pc (corresponds to the knee energy $3 \times 10^6$ GeV in the particle energy spectrum at energies).

Taking into account the turbulent diffusion of CR particles in the diffusion equation, the Laplace operator is considered in degree $\alpha/2$, i.e., a replacement $(-\Delta) \to (-\Delta)^{\alpha/2}$ is made during the transition from normal to anomalous or turbulent diffusion. Therefore, for the resulting exponent $\gamma_p$ of the differential spectrum of proton energy $I(E) = dN(E)/dE \propto E^{-\gamma_p}$, a well-known ratio is obtained (Erlykin & Wolfendale 2001; Lagutin et al. 2001):

$$\gamma_p = \gamma_{\rm ing} + \frac{\alpha}{2}. \tag{14}$$

As a result, from Equations (13) and (14) we obtain the ratio

$$\gamma_p = 3 - \chi + \frac{\alpha}{2}. \tag{15}$$

Thus, at a sufficient distance from the SNR, when it is necessary to take into account the contribution of turbulent diffusion to the propagation of CR particles from observational data in the energy range $10\,{\rm GeV} \leqslant E \leqslant 200\,{\rm GeV}$ in Equation (15), the magnitude $\gamma_p \approx 2.85$ of the exponent can be used.

Suppose that in this region turbulence is still characterized by the spectrum in Equation (12) with the exponent $\chi = 8/3$. Then, to characterize the anomalous turbulent diffusion in turbulence from Equation (15), we obtain an estimate $\alpha \approx 5$. At the same time in the case of the turbulence energy spectrum with an index of $-8/3$, as in Equation (12), we obtain an estimate for the value of the exponent $\gamma_{\rm ing} = 1/3$ from Equation (13). This value is more than 6 times less than the known value $\gamma_{\rm ing} \approx 2.15$ (Erlykin & Wolfendale 2002), which characterizes the direct effect of SNR on the differential energy spectrum of protons escaping into the halo after injection from SNR. Such a large difference indicates that in the considered area not only the turbulence energy is determined by the CR particles themselves (without any connection with their primary source from the SNR), but also the acceleration of CR particles is realized due to some new mechanisms. In particular, in connection with such interaction with ISM, in which not only the emission of the Cherenkov photon by the medium becomes energetically advantageous, but also the additional acceleration of CR particles (Chefranov 2012).

There is also standard relation (Erlykin & Wolfendale 1998) for the diffusion coefficient $D(k) \propto \lambda c/kE_T(k) \propto k^{-(2-\chi)}$, where $\lambda$ is the Larmor radius of the particle of energy $E$ and $E_T(k)$ is the power in irregularities of magnetic field on scale $k = 2\pi/\lambda \propto E^{-1}$.

A variety of forms have been given for the turbulence energy spectra $E_T(k)$ and $D(E)$, depending on how the energy is distributed in the ISM: $D(k) \propto k^{-1/3} \cong O(E^{1/3})$ when the spectrum is $E_T(k) \propto k^{-5/3}$ (Kolmogorov–Obukhov), $D(k) \propto k^{-1/2} \cong O(E^{1/2})$ for the case $E_T(k) \propto k^{-3/2}$ (Kraichnan) and $D(k) \propto k^{-5/3} \cong O(E^{5/3})$ when $E_T(k) \propto k^{-1/3}$ (Gorbatskij; Erlykin & Wolfendale 1998). For the case of spectrum $-8/3$ in Equation (12) the same type of estimation gives $D(k) \propto k^{2/3} \cong O(E^{-2/3})$.

Thus, for the magnetic energy spectrum of compressible turbulence in a medium of CR particles, the turbulent diffusion coefficient decreases with increasing particle energy and, accordingly, its influence on the formation of the differential spectrum of CR energy at sufficiently high energies decreases. On the other hand, it is possible, by analogy with Equation (13), to consider the dependence of the form $\gamma_p = \gamma_{\rm ing} + \delta$, where the designation for the exponent characterizing the dependence $D(E) \propto E^{\delta}$ of the turbulent diffusion coefficient on energy is taken into account. At the same time, for turbulent diffusion in the case of a turbulent energy spectrum of Equation (12) with an index of $-8/3$, as shown above, the value $\delta = -2/3 \approx -0.67$.

If we assume that the value of the exponent $\gamma_p \approx 2.67$, observed in the energy range $200\,{\rm GeV} < E < 3 \times 10^6\,{\rm GeV}$, then we get an estimate $\gamma_{\rm ing} \approx 3.34$, which is close to the observed value $\gamma_p$ for energies $E > 3 \times 10^6$ GeV above the knee, where the turbulent regime with an index of $-8/3$ is no longer present and does not create turbulent diffusion that ensures the delay of CR particles in the inertial scale interval. This possibility has not been considered before, since it is usually assumed that diffusion leads to a relative decrease in the number of particles in a given scale range. This provides an alternative explanation to the conclusions of the work (Erlykin & Wolfendale 1998) of the mechanism of the occurrence of a fracture with a CR energy spectrum called a knee (see Figure 1). Indeed, the compressible turbulence with a characteristic spectrum of $-8/3$ was not considered in Erlykin & Wolfendale (1998), and for the turbulent regimes given there, the positive value $\delta > 0$ of the exponent for the turbulent diffusion coefficient provides, on the contrary, a steeper decrease in the energy spectrum as it increases. Moreover, it can be concluded that only thanks to the implementation of the turbulent scaling regime with an index of $-8/3$, a mechanism for the delay of CR particles in the inertial subrange of scales is provided, which, apparently, is realized only in the energy range 200 GeV $< E < 3 \times 10^6$ GeV.

Indeed, the differential spectrum of proton energy has a noticeably steeper decline with energy growth in the energy range 10 GeV $\leqslant E \leqslant$ 200 GeV, as shown by the results of various observations (Blasi et al. 2012; Neronov et al. 2012) than in the energy range 200 GeV $< E < 3 \times 10^6$ GeV.

The ratio of the fluxes of secondary and primary nuclei in CRs has long been used as an indicator of the grammage traversed in the journey of CR particles throughout the Galaxy (Luo et al. 2023; Ma et al. 2023; Cowsik & Huta 2022; Luoni et al. 2021; Brennan & Coyne 1964; Cole & Cherkaoni-Tadili 1987). At sufficiently high energy, typically 100 GeV/n, the ratio of fluxes of the secondary nucleus to that of the main primary nucleus is found to scale as $E^{-\delta}$, where $E$ is the energy per nucleon (a conserved quantity in spallation reactions) and $\delta$ identifies usually as the energy dependence of the diffusion coefficient (De La Torre Luque et al. 2021). The same shock waves that may be responsible for CR acceleration in the first place also pick up any other charged particle in the upstream, provided being above the threshold for injection. The secondary nuclei produced by spallation in the ISM are no exception, hence they also get accelerated. This effect is unavoidable, only its strength may be the subject of debate. For example, the spectrum of secondary elements such as boron and lithium may be computed by taking into account shock reacceleration (Thoudam & Horandel 2014; Blasi 2017).

Based on the PAMELA and later observations, various options for the formation of turbulence and its effect on the observed change in the energy spectrum of CRs (see Table 1 below from Niu 2021), leading precisely to an exponent close to $-8/3$ is obtained (Blasi et al. 2012; Neronov et al. 2012; Reichherzer et al. 2022).

At the same time, as can be seen from Table 1, it follows that not only near the spectral knee and ankle, corresponding to sharp changes in the slope of the spectrum in Figure 1, but also at the energies indicated in the last column, there is a change in the magnitudes of the power-law exponents, both for primary CR particles and secondary particles. It follows from Figure 1 that data on secondary particles are available only in a relatively limited range of particle energies (stiffness). Therefore, the ratio of secondary particle fluxes to primary particles usually considered may depend on energy not only due to the presence of such a dependence on the diffusion coefficient, but also on what proportion of primary particles from their total number takes part in the birth of secondary particles. As noted in the Introduction, the spectrum for antiprotons turns out to be higher than obtained in the framework of conventional CR diffusion models (Tan & Ng 1983; Di Bernardo et al. 2010; Tomassetti 2012; Vladimirov et al. 2012; Aguilar et al. 2016, 2020; Schroer et al. 2021). We show that for a model of turbulent diffusion following from a turbulent energy spectrum with an index of $-8/3$, the observational data of the spectra of secondary particles, including antiprotons, already admit an acceptable description, if we additionally consider the effect of reacceleration.

**Table 1**
Energy Spectrum Breaking for the Primary and Secondary Particles

| Particles (Atoms) | Before Breaking | After Breaking | Energy (GeV/n) |
|---|---|---|---|
| Proton | −2.815 | −2.71 | 379 [300; 544] |
| Helium | −2.725 | −2.62 | 331 [281; 448] |
| Carbon | −2.74 | −2.64 | 202 [148; 299] |
| Oxygen | −2.696 | −2.49 | 664 [488; 964] |
| Lithium | −3.18 | −2.98 | 123 [112; 351] |
| Beryllium | −3.13 | −2.95 | 199 [173; 438] |
| Boron | −3.10 | −2.84 | 275 [194; 422] |
| Nitrogen | −2.93 | −2.66 | 208 [188; 454] |

Let us consider a new variant of the mechanism of reacceleration of CR particles, in which shock waves accelerating particles are generated in a finite time (Equation (11)) when a simple wave describing the inertia motion of the CR stream collapses or breaks. At the same time, for the resulting indices of the energy spectra of primary and secondary CR particles, we have the following representations:

$$\gamma_{Pr} = s_{Pr} - 2 + \delta \tag{16}$$

$$\gamma_{\rm Sc} = s_{\rm Sc} - 2 + \delta. \tag{17}$$

As noted above from the observational data for protons and antiprotons at energies above 60 GeV (Aguilar et al. 2016), the left parts in Equations (16) and (17) should coincide. At the same time, for energy above 150 GeV, the values of these exponents are close to the value of 8/3. As a result, from Equations (16) and (17), with the diffusion coefficient index $\delta = -2/3$ obtained for the turbulence energy spectrum $-8/3$, it follows that the values, of the power-law exponents characterizing the reacceleration have a value close to the value $s = 5$ given in Equation (5).

From the data in Table 1, for primary and secondary particles from Equations (16) and (17), at the same value of the diffusion index $\delta = -2/3$, different values of the power-law exponents characterizing the reacceleration of primary and secondary CR particles can already be obtained. For example, for the boron atoms power-law exponent has the value $s_{\rm Sc} \approx 5.51$, and the value is $s_{\rm Pr}$ for the carbon atoms. It should be noted that in the work (Blasi 2017), an index $s = 6$, even exceeding this value was noted in connection with the reacceleration mechanism, which corresponds to even weaker compressions in the shock wave according to Equation (1). Indeed, according to Equation (1), for example, a compression value $\kappa \approx 2.3$, is obtained for a value $s = 16/3$ that is less than the estimate $\kappa \approx 2.8$ given in the Introduction, which does not consider the effect of diffusion on the energy spectrum.

Thus, it is shown that taking into account the turbulent diffusion corresponding to the turbulent regime with a scaling index of $-8/3$ establishes the possibility of explaining the mechanism of the implementation of fractures in the observed energy spectrum of CR particles.

In addition, the well-known observations of intense bursts of high-energy gamma rays require special consideration,

understanding the mechanism of which is also important in connection with the problem of the origin of CRs (Dar & De Rujula 2008; Dado & Dar 2012; Amelino-Camelia et al. 1998; Gagero et al. 2017). At the same time, it is possible to interpret these data based on conclusions from the cosmological model that allows for the permanent arising of the ultralight scalar bosons (which then transform into gamma radiation in the presence of a magnetic field) as a result of vacuum polarization during the observed accelerated expansion of the universe (Chefranov & Novikov 2010; Novikov & Chefranov 2011; Mirzoeva & Chefranov 2018). Accordingly, the currently existing ideas about the secondary nature of the observed diffuse gamma radiation having different intensities near and far from the center of the Galaxy must be also modified. As a result, the conclusions following from this difference in the appearance of the CRs energy spectrum depending on the distance from the center of the Galaxy must be also changing.

However, even without taking into account the noted cosmological mechanism of vacuum polarization and the corresponding origin of CRs and gamma rays, as shown above, even a new version of the reacceleration effect itself can contribute to the preservation of a universal energy spectrum $-8/3$ in a wide energy range, which is consistent with the observational data shown in Figure 1. At the same time, the difference from the specified spectral law in the energy region between the knee and the ankle indicates the possibility of the existence of an additional source of particles due to relatively higher values of particle fluxes in this energy region.

We also note the difference between the mechanism of influence of turbulent diffusion on the formation of the CRs energy spectrum considered in this paper and the one proposed in Blasi et al. (2012) to describe the fracture of the CRs energy spectrum near $E = E_{\rm th} \approx 200\,{\rm GeV}/n$, when, during the transition from smaller values to larger values of energy, the steepness of the decay of the spectrum decreases with increasing energy. In Blasi et al. (2012) it is assumed that for energies $E < E_{\rm th} \approx 200\,{\rm GeV}/n$ the diffusion coefficient value is determined by turbulence initiated by sufficiently fast CR particles themselves in the environment, and for energies $E > E_{\rm th} \approx 200\,{\rm GeV}/n$ the propagation of CRs is already regulated by external turbulence existing before the arrival of CRs with the Kolmogorov–Obukhov energy spectrum $-5/3$.

In our paper, on the contrary, taking into account the dominance of the $-8/3$ turbulent spectrum in the inertial range observed according to Figure 2, it is assumed that for energies $E < E_{\rm th} \approx 200\,{\rm GeV/n}$ one or another mechanism of the occurrence of external turbulence with the Kolmogorov–Obukhov spectrum $-5/3$ is realized.

And only when $E > E_{\rm th} \approx 200\,{\rm GeV}/n$ in the medium itself consisting only of CR particles, self-turbulence arises with a spectrum of $-8/3$ and a corresponding coefficient of turbulent diffusion $D(E) \propto E^{-2/3}$, which leads to a steepening of the spectrum in this energy region according to the observational data shown in Figure 1.

## 5. Conclusions

In this paper, the isothermal mode of propagation of CR particles corresponding to the conservation of their energy is considered. This is described in the framework of the well-known two-liquid model (Malkov & O'C Drury 2001) of the interaction of the medium of CR particles and the medium in which they propagate. In this case, a closed description of the dynamics of the CR medium is carried out on the basis of one-dimensional Euler equations for a compressible medium. Taking into account the exact explicit solution of these equations (Chefranov & Chefranov 2021), an explicit analytical form of a two-point correlation function for the CR particle density and to the corresponding universal spectrum $-8/3$ for CR particle density fluctuations in the inertial subrange of scales is obtained, which is matched with known observation data (Armstrong et al. 1995; see also Figure 1). In contrast to the consideration of the two-fluid model of CR particle dynamics in Malkov & O'C Drury (2001), due to the use of the isothermal regime obtained in Skilling (1971, 1975), the Euler equations for the CR medium are uncoupled from the equations describing the medium in which CRs propagate. This made it possible to find and use the corresponding exact solution of the Euler equations to describe the spectrum of fluctuations in the density of CR particles, with the universal scaling law $-8/3$.

It is necessary to carry out additional measurements of two-point correlations for fluctuations in the density of CR particles and the corresponding spectrum as in Figure 2, based on the already available data on fluctuations in the density of CR particles measured in the PAMELA experiment.

Indeed, at present, for the differential energy spectrum and the corresponding specific density $N(t;\ \vec{x};\ \vec{p})$ of CR particles, observational data are known that allow us to establish a power-law dependence of the type $N \propto p^{-\gamma}$ at the measurement point with an exponent close to $-8/3$ and sometimes reaching a value of $-3$, which is consistent with theory (Skilling 1975) and present consideration. At the same time, however, the proportionality coefficient in this scaling law can be associated with the total density $n(\vec{x};\ \ t)$ fluctuating in space and time, the two-point correlation function of which corresponds to the obtained exact turbulent spectrum of Equation (12).

This will make it possible to establish the mechanism of formation of such a spectrum, based on comparison with the exact solution obtained in this paper when determining the scaling exponent $-8/3$ for the turbulent spectrum of fluctuations in the density CR particles, based only on the exact solution of the Euler equations for a compressible medium. Of interest is also the possibility of establishing a correlation between the scaling law $-8/3$ observed in the turbulent spectra of the cosmic plasma in the solar wind and magnetosheath (Passot & Sulem 2015; Chefranov & Chefranov 2021) and the similar exponent of the turbulent spectrum of fluctuations in the density of particles of CRs obtained in this work, and also with a similar spectrum of $-8/3$, characteristic of fluctuations in the intensity of the surface earth electric field (Anisimov et al. 2002). It is of interest to further use the obtained exact solutions in the theory of turbulence and in the new theory of Vavilov–Cherenkov radiation (Chefranov 2004, 2012, 2016) to analyze the problems of describing such turbulence and to consider the mechanism of acceleration of CR particles and their origin. In our paper, it is also shown that shock waves can spontaneously form in a medium consisting of CR particles, which, like shock waves arising from the explosion of supernovae, can cause additional particle acceleration. At the same time, for the corresponding nonequilibrium system, the characteristic scale of which is commensurate with the thickness of the shock wave, it is possible to deviate from the statistically stationary regime associated with the second law of thermodynamics, as is the case for the atmosphere (Singh & O'Neill 2022). Also, a vortex can accumulate energy and can attain higher intensity than previously existed or is observed in its neighborhood. It gives an example of the so-called animated growth of vortical systems (Fujiwhara 1923) which is a kind of selection phenomenon in analogy with an example of selection like mechanical or electromagnetic resonance.

Thus, the application of an exact solution to the problem of one-dimensional compressible turbulence indicates additional new possibilities for using the theory of hydrodynamic turbulence for various nonequilibrium systems (Chefranov & Chefranov 2021, 2022).

## Data Availability

The data that support the findings of this study are available from the corresponding author upon reasonable request.